# A Letter to the NSF Astronomy Portfolio Review: NOAO 4-m Telescopes for Future Surveys

## David Schlegel (LBNL), Mike Sholl (UC Berkeley SSL)
## 31 January 2012

The NOAO currently operates the two most capable platforms in the world for optical surveys, the 4-m KPNO (Mayall) and 4-m CTIO (Blanco) telescopes. It was only discovered recently (in 2009) that a field of view of 3 deg diameter is possible on these telescopes. In combination, these two telescopes provide the unique capability of a common telescope platform for full-sky surveys. The survey power (in étendue) is 45% that of LSST.

Two ambitious surveys are already planned using these telescopes in the coming decade: Dark Energy Survey (imaging on the Blanco) and BigBOSS (spectroscopy on the Mayall). The BigBOSS collaboration has proposed a survey of 20 million galaxies in one hemisphere, and the full sky could be completed by moving the instrument to its sister platform in the South. These and other possible surveys argue for the continued investment in these uniquely-capable facilities by NSF Astronomy.

**Mayall and Blanco Wide Field Designs**

The Mayall and Blanco telescopes were designed in the 1960s and commissioned in 1973 and 1974. The Ritchey–Chrétien designs allow long focal lengths and large fields of view. The current secondary mirrors support instruments with fields-of-view up to 36 arcmin. Newer optical designs have shattered this limit: The Dark Energy Camera includes a prime focus corrector for a 2.2 deg diameter field, and the BigBOSS spectrograph includes a prime focus corrector for a 3.0 deg diameter field. Other optical solutions are possible at either the prime or Cassegrain configuration with fields of view larger than 3 deg (http://www-kpno.kpno.noao.edu/kpno-misc/mayall_params.html).

**A Survey of Wide-Area Telescope Platforms**

A survey of all 4-m-class telescopes was conducted by one of the authors (M. Sholl) in 2009-2011 to establish their viability for wide FOV imaging and spectroscopy. This includes all 17 known astronomical and military telescopes (see Table 1). Not included in this study were the 3.5-m telescope at Mount Korek, Iraq, for which the optical design was not available, and the UKIRT 4-m.

A handful of these telescopes are suitable for a 3 deg diameter field. The best optical solutions are found for the Mayall, Blanco, and CFHT. Solutions are possible on AAT and Calar Alto, although the image quality on these sites compromises their suitability for future surveys of faint objects. (They might represent opportunities for high-resolution spectroscopic surveys of stars.)

AEOS is a military telescope, and NTT ESO and ESO 3.6m are committed for the foreseeable future with modern instruments.

The Mayall and Blanco telescopes have the additional advantage that the primary mirror figures are nearly identical. An optical corrector for one telescope (including the secondary mirror for a Cassegrain design) can be used on its sister telescope with a re-spacing of the lenses in their cells. This is the ***only*** pair of 4-m telescopes for which this is possible, and provides a very cost-effective opportunity for full-sky surveys using a common instrument.

Most modern 4-m telescope designs favor much faster optics (smaller f/#'s). These designs are much more compact physically, resulting in smaller and more cost-effective telescopes and enclosures. It is interesting to note that the most recent 4-m telescope to be commissioned, VISTA at Cerro Paranal, has the fastest optics at f/1.0. The drawback to these designs is that the larger focal planes are not possible due to the fast optical speed.

The Mayall and Blanco telescopes can support large instruments at either prime or Cassegrain focus. By replacing the f/8 flip ring, it is possible to support the weight of four Toyota Land Cruisers (5800 lb. curb weight each). This easily supports instruments such as the BigBOSS optics and spectrograph, or perhaps even more ambitious instruments in the future.

**Comparison to SDSS and LSST**

The Mayall and Blanco telescopes are superior to any existing or planned survey focal planes except Subaru/HSC and LSST. The traditional measure of survey speed is the étendue, defined as the product of the effective collecting area and field of view. Each of these telescopes has an étendue 2.5 times larger than SDSS and 4.4 times larger than PanSTARRS-1 (c.f., Table 2).

LSST represents the most capability for a large area imaging survey once it is constructed. The LSST focal plane is unlikely to be re-purposed for spectroscopy as its f/1.2 focal length is a poor match to slit or fiber spectroscopic instruments. Subaru/HSC is planned to be used for a wide-field spectrograph, but that time will be split between many other instruments and will be available exclusively to the Japanese community. The Mayall and Blanco platforms, with a combined étendue of 45% of LSST, represent the best spectroscopic survey capability in the future. As survey platforms, these are unique resources for U.S. astronomy.

| Name | Site | Notes and Exclusions | M1 f/# | M1 Diam. (m) | f/# | f (m) | Suitable for BigBOSS corrector? |
|---|---|---|---|---|---|---|---|
| Vista | Cerro Paranal, Chile | ESO Committed | f/1.0 | 4.1 | 1 | 4.1 | no |
| Starfire | Kirtland AFB, New Mexico | Military | f/1.5 | 3.5 | 1.5 | 5.25 | no |
| SOAR | Cerro Pachon, Chile | | f/1.7 | 4.2 | 1.7 | 7.14 | no |
| WIYN | Kitt Peak, Arizona | | f/1.8 | 3.5 | 1.8 | 6.3 | no |
| ARC | Apache Point, New Mexico | | f/1.8 | 3.5 | 1.8 | 6.3 | no |
| Discovery Channel | Lowell Obs, Arizona | | f/1.9 | 4.2 | 1.9 | 7.98 | marginal, with 1.5m C1 |
| Galileo TNG | La Palma, Canary Islands, Spain | | f/2.2 | 3.6 | 2.2 | 7.92 | marginal |
| NTT ESO | Cerro La Silla, Chile | ESO Committed | f/2.2 | 3.5 | 2.2 | 7.7 | yes |
| William Herschel | La Palma, Canary Islands, Spain | | f/2.5 | 4.2 | 2.5 | 10.5 | marginal |
| Victor Blanco | Cerro Tololo, Chile | Twin to Mayall | f/2.8 | 4 | 2.8 | 11.2 | yes |
| Mayall | Kitt Peak, Arizona | Twin to Blanco | f/2.8 | 3.8 | 2.8 | 10.64 | yes |
| AEOS | Maui, Hawaii | Military | f/3.0 | 3.7 | 3 | 11.1 | yes |
| ESO 3.6m | Cerro La Silla, Chile | ESO Committed | f/3.0 | 3.6 | 3 | 10.8 | yes |
| AAT | Coonabarabran, NSW, Australia | 2 arcsec seeing | f/3.2 | 3.9 | 3.22 | 12.558 | yes |
| Hale | Palomar Mountain, California | | f/3.3 | 5.1 | 3.3 | 16.83 | no, massive corrector |
| MPI-CAHA | Calar Alto, Spain | Poor seeing | f/3.5 | 3.5 | 3.5 | 12.25 | yes |
| CFHT | Mauna Kea, Hawaii | Proposed 10m | f/3.8 | 3.6 | 3.8 | 13.68 | yes |

Table 1: Result of optical study of 17 telescope for the 3 deg FOV BigBOSS corrector, sorted from fastest to slowest optics.

| Telescope | Primary mirror | Central obscuration | Mirror losses | FOV | Etendue (Area * FOV) |
|---|---|---|---|---|---|
| SDSS | 2.5 m | 1.25 m | $(0.9)^2$ | 7.0 deg$^2$ | 20.9 m$^2$ deg$^2$ |
| PanSTARRS-1 | 1.8 m | 0.9 m | 0.9 | 7.0 deg$^2$ | 12.0 m$^2$ deg$^2$ |
| Mayall/BigBOSS | 3.8 m | 1.8 m | 0.9 | 7.0 deg$^2$ | 55.4 m$^2$ deg$^2$ |
| Blanco/DECam | 4.0 m | 1.6 m | 0.9 | 3.0 deg$^2$ | 28.5 m$^2$ deg$^2$ |
| LSST | 8.4 m | 5.06 m | $(0.9)^3$ | 9.6 deg$^2$ | 247 m$^2$ deg$^2$ |
| Keck | 10.0 m | 1.4 m | $(0.9)^3$ | 0.087 deg$^2$ | 4.88 m$^2$ deg$^2$ |
| Subaru/HSC | 8.2 m | 0.95 m | 0.9 | 1.8 deg$^2$ | 84.4 m$^2$ deg$^2$ |

Table 2: Comparison of survey speed (étendue) possible on existing (SDSS, Blanco) and future (Mayall, LSST) platforms. Almost every other larger telescope (such as Keck) are not appropriate for large-area surveys owing to their low étendue, with the exception of Subaru/HSC.